\documentclass[footinbib,twocolumn,showpacs,amsmath,amstex,amssymb,mathfonts,superscriptaddress,prx]{revtex4}
\usepackage{graphicx}
\usepackage{color}
\usepackage{bm}

\usepackage{graphicx}
\usepackage{xcolor}
\usepackage{bm}
\usepackage{amsmath}
\usepackage{amssymb}
\usepackage{amsthm}
\usepackage{amsfonts}
\usepackage{multirow}
\usepackage{makecell}
\usepackage{float}
\usepackage{comment}
\usepackage{enumitem}
\usepackage{verbatim}
\usepackage{mathtools}

\begin{document}

\title{Edge modes of topologically ordered systems as emergent integrable flows: Robustness of algebraic structures in nonlinear quantum fluid dynamics}
\author{Yoshiki Fukusumi}
\affiliation{Center for Theory and Computation, National Tsing Hua University, Hsinchu 300044, Taiwan}
\affiliation{Physics Division, National Center of Theoretical Sciences, National Taiwan University, Taipei 106319, Taiwan}

\date{\today}
\begin{abstract}

In these decades, it has been gradually established that edge modes of a wide class of topologically ordered systems are governed by the bulk-edge correspondence and anyon condensation. The former has been studied many times because it can be summarized as the correspondence between conformal field theory and topological quantum field theory, but the latter has captured attention more recently because it suggests that one can deform the theories from one to another condensable or Witt equivalent.  We reinterpret this phenomenon as the appearance of integrable flow at the edges of a topologically ordered system. By revisiting the existing phenomena from this view, one can relate the anyon condensation to the existing formalism of integrable models and their renormalization group flow. Moreover, we observe the robustness of the bulk-edge correspondence even under the existence of irrelevant perturbations at the edges. This formulation can enable one to analyze edge modes at the level of quantum fluid (non-linear) dynamics, or mixed state topological order even in non-Hermitian systems. Moreover, especially in the study of a class of non-Hermitian systems, we show a fundamental necessity to revisit Rogers-Ramanujan type identities as an appearance of the hidden (fractional) supersymmetric model which has a close relation to string field theories. Our view can give a unified paradigm to study edge modes of topologically ordered systems at the level of their dynamics potentially in both Hermitian and non-Hermitian systems. 
\end{abstract}

\maketitle

\section{introduction}

Anyons and their condensations\cite{Kong:2013aya,Burnell_2018} are a key structure to construct and analyze topological order (TO) and quantum phase transitions\cite{Thouless:1982zz,1985AnPhy.160..343K,Wen:1989zg,Wen:1989iv}. Their history is very long if one includes the studies of massless flow\cite{Zamolodchikov:1989hfa} and (hidden) symmetry classification of spin chains\cite{Haldane:1982rj,Haldane:1983ru,haldane2016groundstatepropertiesantiferromagnetic,Kennedy:1992ifl,MOshikawa_1992}, known as symmetry-protected topological (SPT)  phase in modern terminology\cite{Pollmann_2010,Chen:2010zpc}. More recently, by applying mathematical frameworks to these studies such as the modular tensor category (MTC), many series of TOs have been constructed and analyzed\footnote{There exist a large number of references on this subject with various formalisms, so we list  recent reviews and a textbook \cite{Wen:2015qgn,Simon:2023hdq,Kong:2022cpy} which can be helpful for the readers}. Roughly speaking, this kind of studies are focusing on the algebraic aspects of TOs or corresponding topological quantum field theories (TQFTs) with a close connection to opeartor algebraic structure of the conformal field theories (CFTs). This correspondence is known as bulk-edge correspondence or CFT/TQFT correspondence\cite{Witten:1988hf}, and appears ubiqutously in contemporary physics\cite{moore_nonabelions_1991}.

Hence, there may arise a natural question from a wider audience on the necessity of further theoretical studies of TOs. Moreover, theoretical analysis of the dynamical property of the fractional quantum (FQH) fluid \cite{Abanov:2005nt,Wiegmann_2012,Nardin_2020,Nardin:2022mto,nardin2023refermionizedtheoryedgemodes,Nardin:2023byl} has not captured sufficient attention in the fields regardless of their fundamental importance in the numerical and experimental settings. Hence one may think of a CFT as a knowledgeable theory with fruitful predictions, but not as a tool to calculate or construct the relevant observables in the experimental and numerical settings.

In this work, we propose a modern interpretation of edge modes of TOs based on recent theoretical (or mathematical, to some extent) studies. We interpret the edge modes as \emph{emergent integrable flow}\footnote{The terminology, \emph{emergent} has been used roughly in the field, but we use this terminology from a traditional view: the object that is difficult to stabilize in a ultra-violet (UV) theory (or a microscopic model) but is stabilized in the infrared (IR) regime under the renormalization group\cite{Anderson:1972pca}.}. This gives a unification of various existing formalisms and explains various phenomena in the fields, and it is useful to obtain a unified understanding and build a strategy to calculate observables becuase of the integrability. Moreover, we point out the fundamental importance of studying false modular forms\cite{Bringmann:2019vyd,Bringmann:2021dxg} for the quantitative and qualitative understanding of (nonlinear) dynamics of quantum fluids, kown as the quantum Korteweg-de Vries equation (KdV) \cite{Sasaki:1987mm,Kupershmidt:1989bf,Eguchi:1989hs,Bazhanov:1994ft} for the edge of $2+1$ dimensional TOs. At this stage, one can see the importance of our formalism and underlying CFTs as a possible strategy to construct TOs in a quantitative (or mathematical) language. 

It should also be stressed that our strategy needs only the bulk-edge correspondence and their renormalization group (RG) interpretations. Hence, one can apply it to non-Hermitian analogs in principle. Following the studies on the level-rank duality \cite{Nakanishi:1990hj,Naculich:1990hg,Seiberg:2016gmd,Hsin:2016blu,Cordova:2017vab} and recent development of stuides on $3d-3d$ correspondence\cite{Gukov:2017kmk,Cho:2022kzf,Gang:2023ggt,Gang:2023rei,Baek:2024tuo,Gang:2024tlp,Creutzig:2024ljv}, we propose a concise understanding of a class of general TOs. In this background, surprising for condensed matter physicists, the studies on $Osp(m,2n)$ Wess-Zumino-Witten (WZW) models and related (fractional) supersymmetric models are fundamentally important\cite{Saleur:1989gj,Bershadsky:1989tc,Saleur:2001cw,Saleur:2003zm,Yang:2008vb}. These models have been revisited to study $3d$ invariant\cite{Cho:2022kzf,Gang:2023ggt,Gang:2023rei,Baek:2024tuo,Gang:2024tlp,Creutzig:2024ljv}, but we propose their importance also in the study of non-Hermitian models. One can even see literature on M-theory \cite{Gunaydin:1998bt,Park:2005pz}, but this might be reasonable (not surprising) when considering the close connection between Kitaev's $16$-fold way and fermionic string theories\cite{Kitaev:2006lla}. It is also worth mentioning that the appearance of Jack polynomial in the studies of fractional quantum Hall systems  (or the most established TOs) \cite{Bernevig2008PropertiesON,Bernevig_2008} which has a connection to Rogers-Ramanujan type identities and fermionic or supersymmetric models\cite{feigin2001differentialidealsymmetricpolynomials}\footnote{Corresponding integrable flows of nonunitary models have been studied in \cite{Tanaka:2024igj} very recently.}.

The rest of the manuscript is organized as follows. In Section \ref{main_claim}, we propose a unified understanding of edge modes of topological order as a combination of Witt equivalent (anyon condensation) and integrable deformation by irrelevant operators (or quantum KdV hierarchy). The readers can see the fundamental importance of the study of false modular forms in mathematics and mathmatical physics. In Section \ref{non-Hermitian_correspondence}, we apply our formalism to more nontrivial situations, TOs of non-Hermitian systems. This section contains interpretations of a surprising equivalence between a nonunitary CFT and the unitary CFT which corresponds to similarity transformation in the many-body systems. We clarify that this phenomenon can be understood as hidden supersymmetry and its close analogy with the dimensional reduction in the literature. Section \ref{conclusion} is the concluding remark of this work with mention on possible future directions and open problems.

\emph{Notes for the readers}: In this work,we have tried as respectful as possible to various research fields by making many historical remarks. The main thoretical views have been summarized in figures in the manuscript, so we expect that the readers can understand the main concepts by checking the figures and corresponding arguments in the main text.

\section{Emergent integrable flow}
\label{main_claim}
\begin{figure}[htbp]
\begin{center}
\includegraphics[width=0.5\textwidth]{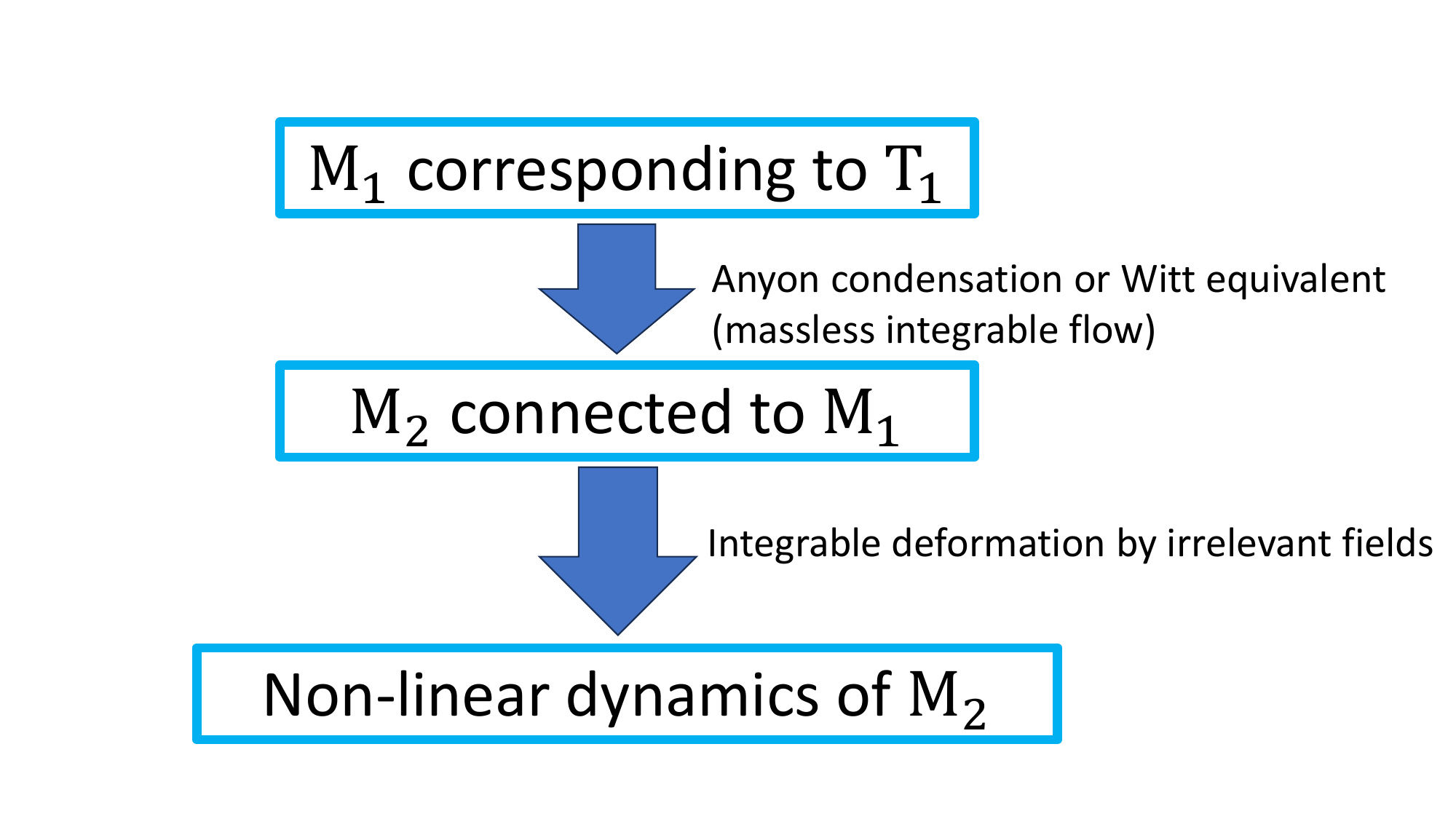}
\caption{A unified understanding toward descriptions of algebraic and dynamical properties of emergent edge modes. One can study the algebraic property or fusion rule of anyons in constructing the top arrow, and the dynamical property in the bottom arrow. As we discuss, recent studies indicate the algebraic properties are preserved in the bottom arrow. }
\label{robustness_of_integrability}
\end{center}
\end{figure}

In this section, we propose the main conjecture or unified interpretation of edge modes appearing in a TO (FIG. \ref{robustness_of_integrability}). First let us assume the bulk TO $\mathbf{T}_{1}$ corresponds to an CFT $\mathbf{M}_{1}$. Then we propose the following theory described by a Hamiltonian $H_{\text{b}}$ should appear at the boundary (or  $H_{\text{b}}$ should give a class of computable series of theories),

\begin{equation}
H_{\text{b}}=H_{\mathbf{M}_{2}}+\sum_{i} \lambda_{i} I_{i} 
\label{Witt_Gibbs}
\end{equation}
where $\mathbf{M}_{2}$ is an CFT corresponding to TO $\mathbf{T}_{2}$ which is Witt equivalent to $\mathbf{T}_{1}$\cite{Kaidi:2021gbs,Fukusumi_2022_c,zhang2024hierarchyconstructionnonabelianfractional}, $H_{\mathbf{M}_{2}}$ is the Hamiltonian of  $\mathbf{M}_{2}$ and $I_{i}$ and $\lambda_{i}$ are higher local integrals of motion in $H_{\mathbf{M}_{2}}$and their coupling constants. $I_{i}$ appears in the study of quantum  Korteweg-de Vries (KdV) hierarchy \cite{Sasaki:1987mm,Kupershmidt:1989bf,Eguchi:1989hs,Bazhanov:1994ft}, a class of quantum fluid governed by non-linear equations. By taking a trace, this corresponds to a generalized Gibbs ensemble \cite{Novaes:2021vjh} and one can naturally understand its relation to TO including mixed state. However, their implication is much stronger than naive expectations as we will demonstrate. Related studies on the nonlinear dynamics of fractional quantum Hall fluid can be seen in  \cite{Abanov:2005nt,Wiegmann_2012,Nardin_2020,Nardin:2022mto,nardin2023refermionizedtheoryedgemodes,Nardin:2023byl,Novaes:2021vjh} and our proposal is a theoretical unification of these studies. More intuitively, it is known that there exists geometric excitation in the FQHE liquid\cite{Haldane:2011ia,Yang_2012,Yuzhu_2023}. The existence of such excitations corresponds to the weak breaking of conformal invariance\cite{Cardy:1986ie} of the edge theories whereas their topological property is preserved. This corresponds to the robustness of the topological property and one can expect the robustness of the fusion rule which should govern the properties of the anyons.

As a summary, based on the detailed interpretation of Eq. \eqref{Witt_Gibbs}, we propose the main claim of this work as follows:
\begin{itemize} 
\item{\emph{To obtain a general understanding of anyon condensations appearing modern theoretical condensed matter, it is fundamentally important to study (or) revisit the structures of integrable flow.}}
\item{\emph{The calculus of false modular characters is fundamentally important to explore the (nonlinear) response theory of quantum fluid appearing at the edge of topologically ordered systems.}}
\end{itemize}

In the following subsection, we explain the interpretations of each term in Eq. \eqref{Witt_Gibbs}. Moreover, in Section \ref{non-Hermitian_correspondence}, we observe the implication of our formalism in the study of recent non-Hermitian systems and the related non-local Hermitian systems.

\subsection{Massless flow and Witt equivalence}
\label{Witt_RG}

In this subsection, we observe the implication of the first term $H_{\mathbf{M}_{2}}$ in Eq. \eqref{Witt_Gibbs}. A relevant thing to note for the audience unfamiliar with the subjects is that the central charge of two theories $\mathbf{M}_{1}$, $\mathbf{M}_{2}$ can be different\cite{Kaidi:2021gbs} and the central charge of the boundary theory $H_{\mathbf{M}_{2}}$ can be larger than that of $\mathbf{M}_{2}$\cite{Fukusumi:2020irh,Fukusumi_2022_c,Fukusumi:2023vjm}. More surprisingly, this explains the controversial thermal Hall conductance in experimental settings\cite{Mross_2018,Wang2017TopologicalOF} without inconsistency at this stage\footnote{This may correspond to a modern version of the discussions in \cite{Son:2015xqa}.}. 

Theoretically, this is closely related to the massless flow of CFTs proposed by Zamolodchikov\cite{Zamolodchikov:1989hfa} whereas this aspect has rarely been discussed in the study of generalized symmetry. In this RG flow, CFTs with different central charges can be deformed smoothly (or without obstruction coming from the mismatch of $Z_{N}$ symmetry or Lieb-Shultz-Mattis type anomaly\cite{Cho:2017fgz,Yao:2018kel,Fukusumi_2022_c}) and one can relate the phenomena to conformal interface or RG domain wall\cite{Gaiotto:2012np}\footnote{The anomaly matching or cancellation condition is consequence of the structure of the integer-spin simple current by Schellekens and Gato-Rivera\cite{Gato-Rivera:1990lxi,Gato-Rivera:1991bqv,Gato-Rivera:1991bcq,Kikuchi:2022ipr,Fukusumi_2022_c}, also known as the discrete torsion\cite{Vafa:1989ih}.}. The corresponding phenomena in $1+1$ dimensional quantum chains and corresponding CFTs have been studied \cite{Lecheminant:2015iga,Furuya:2015coa,Kikuchi:2019ytf,Kikuchi:2022ipr,Lecheminant:2024apm}. Because of CFT/TQFT correspondece\cite{Witten:1988hf}, one can interpret such symmetry-based RG arguments as classification of TOs. In contemporary theoretical physics, this is studied in the name anyon condensation with close connection to ``topological holography"\cite{Moradi:2022lqp,Huang:2023pyk}, and one can relate the phenomena to (fractional) supersymmetric models \cite{Wen:2024udn,Fukusumi:2024cnl,Huang:2024ror,Bhardwaj:2024ydc} \footnote{However, it is worth mentioning that the original terminology of ``topological holography" appeared in a different context to indicate a system both ``topological" and ``holographic" in \cite{Husain:1998vz} as the author has mentioned in\cite{Fukusumi:2024cnl}.}.

\subsection{Quantum integrable hierarchies, modular properties: Robustness of fusion rules}
\label{false_modular}

In a conformal field theory, a series of descendant fields exist. Hence without suitable assumptions, they can appear as perturbations of a model corresponding to a many-body system. However, as is well established in the fields, such irrelevant perturbations break the integrability of the model in general. The remarkable consequence of this breaking of integrability is that the response theory (including the nonlinear one) to the gapless system is broken \cite{shimizu1999nonequilibriummesoscopicconductorsdriven,Fukusumi:2021qwa,Liu:2021gyt,PhysRevB.104.205116} \footnote{Historically, this corresponds to Van Kampen's objection to the response theory \cite{Kampen_1971,VANVELSEN1978135}. For the readers interested in its relation to TOs, we note the concluding remark of \cite{1985AnPhy.160..343K}, and related discussion in the author's works \cite{Fukusumi:2021qwa,Fukusumi:2022xxe}.}. The structure of the Hilbert space of a CFT can be deformed quite nontrivially under such irrelevant perturbations, and operators constructed from such deformed Hilbert space can be different from the original CFT $\mathbf{M}_{2}$. Hence bulk-edge correspondence can easily be broken by such irrelevant perturbations. 

A simplest example is $U(1)$ CFT perturbed by the chiral Umklapp term. For simplicity, let us consider chiral CFT with primary fields $e^{im\varphi}$ with conformal dimension $m^{2}$, where $m$ is an integer corresponding to momentum or $U(1)$ charge and $\varphi$ is a bosonic field\footnote{The arguments do not depend on the compactified radius or level of $U(1)$ Kac-Moody algebra, so let us be a bit sloppy about that point}. By definition,  if one introduces the chiral Umklapp term $\text{cos}n\varphi$, the degeneracy coming from the operation $m\leftrightarrow -m$ can be slightly broken and the set of primary fields become $\text{cos} m_{c} \varphi$ and $\text{sin} m_{s} \varphi$ where $m_{c}$ and $m_{s}$ are positive integers. In this setup, the edge modes are labeled by $m_{c}$ and $m_{s}$. Hence the bulk edge correspondence is broken if there exist nonintegrable perturbations because the wavefunction of the system should be determined by $e^{im\varphi}$. Hence when a system satisfies the bulk-edge correspondence, the form of irrelevant perturbation should be restricted.  

The surprising result of the recent studies on integrable model, especially for quantum KdV Hierarchy, is that the structure of Hilbert space of underlying CFT survives even with integrable perturbations at the level of modular properties of the generalized characters\cite{Downing:2021mfw,Downing:2023lop,Downing:2023lnp,Mulokwe:2024zjc}\footnote{Similar proposal from different methods can be seen in \cite{Ruzza_2021,vanittersum2024quantumkdvhierarchyquasimodular}.}. This aspect was commented on in the concluding remark of \cite{Novaes:2021vjh} for example, but the implications have not been studied fully\footnote{Moreover, really unfortunately, the citation number of \cite{Novaes:2021vjh} is zero at least now.}. First, let us introduce the chiral character of the CFT $\mathcal{M}_{2}$ as $\chi_{a}(\tau)$, where $a$ is the label of primary fields corresponding to anyonic objects and $\tau$ is the modular parameter\footnote{More precisely, the set of the label $a$ depends on the representations (bosonic, fermionic, and so on). For example, the topological symmetry operator of the categorical symmetry, anyons that appear in the construction of the wavefunctions can be different\cite{Fukusumi:2023psx,Fukusumi:2024cnl}. However, they are mutually related by bulk-edge correspondence and Schottky double (or topological holography in the recent literature).}. For later use, we also introduce the parameter $q=e^{2\pi i \tau}$ corresponding to the Boltzmann weight. The following two modular transformations are fundamental,
\begin{equation}
T: \tau \rightarrow \tau +1
\end{equation}
\begin{equation}
S: \tau \rightarrow \frac{-1}{\tau}
\end{equation}
The modular $T$ transformation characterizes the conformal spin of each sector and modular $S$ transformation has been studied in the context of high-low temperature duality\footnote{There exist generalization of modular transformation\cite{Cappelli:1996np}, but to consider fusion rules (or topological properties) of a model, these two transformations are fundamental for our purpose.}. Whereas the interpretation of the modular $T$ transformation is simple, the interpretation of modular $S$ transformation or its intuition is not so at first sight. However, this transformation is important as a fundamental structure of CFTs, because it is closely related to the fusion rule of primary fields by Verlinde formular\cite{Verlinde:1988sn,Cardy:1989ir}. In other words, the algebraic structure of CFTs and the corresponding TOs (including non-abelian) is determined by the modular $S$ transformation (at least in a class of model\cite{Gannon:2001ki}).     

In the recent literature, by modifying the modular $S$ transformation $\hat{\chi}_{a}(-1/\tau) \rightarrow \sum_{b} S_{ab}\tilde{\chi_{b}}(\tau)$, where the $\hat{\chi}$ and $\tilde{\chi}$ are related by complicated transformation for false modular form as in\cite{Downing:2021mfw,Downing:2023lop,Downing:2023lnp,Mulokwe:2024zjc}. Hence one can still observe the modular property of the original CFT without perturbation. This generalized character is called \emph{false modular form} \cite{Bringmann:2021dxg} and studied in the context of $3d-3d$ correspondence\cite{Gukov:2017kmk}. Combining the Verlinde formula with this, the robustness of the modular properties implies that the structure of the fusion rule of primary fields (or nonabelian anyons) in the underlying CFT is also robust against such perturbations. Moreover, a similar proposal at the level of the entanglement spectrum can be seen in \cite{Arildsen_2022} For the readers interested in numerical analysis of related models, we note \cite{liu2024finitesizecorrectionsenergyspectra,Poghosyan:2019yxg}

By applying the correspondence between anyonic objects and generalized symmetry as in \cite{Petkova:2000ip,Fuchs:2002cm,Fuchs:2004dz,Fukusumi:2024cnl}, our formulation ensures the fundamental strength of the analysis of generalized symmetry\cite{Gaiotto:2014kfa}. Because we have studied characters corresponding to the generalized Gibbs ensemble,  one can interpret that the system is in a mixed state\cite{Wang:2023uoj,Sohal:2024qvq,Ellison:2024svg,Kikuchi:2024ibt}. In other words, whereas anyon condensation should give algebraic properties of TOs \cite{Wang:2023uoj,Sohal:2024qvq,Ellison:2024svg,Kikuchi:2024ibt}, the dynamical property is given by generalized Gibbs ensemble and quantum KdV equation\cite{Sasaki:1987mm,Kupershmidt:1989bf,Eguchi:1989hs,Bazhanov:1994ft} without inconsistencies. Because of its close connection to integrability and quantum fluid dynamics, the generalized Gibbs ensemble can be analyzed by nonlinear response theory in principle\cite{10.1143/PTP.15.77,Kubo:1957mj,Watanabe_2020}.

\section{Implications for non-Hermitian (or the corresponding non-local Hermitian) systems: Usefulness of $3d-3d$ correspondence and similarity transformation}
\label{non-Hermitian_correspondence}
Recently, studies of non-unitary conformal field theory and the corresponding nonHermitian or Hermitian lattice or microscopic models have captured the attention of theoretical physicists\footnote{For example see the review \cite{Ashida:2020dkc,Okuma:2022bnb} and references therein. For our purpose, it should be mentioned that the non-Hermitian models have been commonly studied in the context of two-dimensional statistical models and its close connection to quantum group\cite{Pasquier:1989kd} with a connection to black-hole physics\cite{Ikhlef:2011ay}. See discussion in \cite{Korff:2008xx}.}. Moreover, there exist several proposals for realizing non-Hermitian systems corresponding to nonunitary CFTs and TOs in numerical or experimental settings\cite{Lootens:2019xjv,Sanno_2022,Shen:2023tst}. Because of the correspondence between the quantum field theory and lattice models, one can expect a general class of CFTs should essentially give a class of classification of non-Hermitian many-body systems. We note \cite{Lootens:2019xjv} as a recent study of Tensor-Network study of TOs corresponding to the nonunitary CFTs. 

For example, the series of flows between  $M(2, 2N+3)$ minimal models or multicritical Yang-Lee edge models \cite{Nahm:1992sx,Lencses:2022ira,Lencses:2023evr,Lencses:2024wib} have non-abelian anyon and the corresponding structure of topological symmetry or defects. It is also well known that in a class of non-Hermitian models, there exists mapping to the Hermitian model, called \emph{similarity transformation}. In general, the mapping can result in a non-local Hermitian Hamiltonian\cite{Castro-Alvaredo:2009xex}, and the classification of criticality or topological order of non-Hermitian systems is fundamentally important also for the classification of Topological order and criticality with non-local Hermitian Hamiltonian (For the readers interested in the recent studies on non-local models, we note \cite{doi:10.1142/S0217984996000080,Hakobyan_2003,Jones:2022dvn,Gong:2022ufm,Zhou:2024whw,Yang:2024rwz,Ma:2024sig} and appearance of nonlocal interaction under Teo-Touless limit of gapless FQHE\cite{Papi__2014}) \footnote{Related proposal can be seen in the realization of chiral CFT in local non-Hermitian or nonequlibrium systems \cite{Budich:2016zez,Chernodub:2017lmx,Chen:2020kix} or non-local Hermitian systems\cite{Drell:1976mj,Stacey:1981ki,Kaplan:1992bt,,Wang_2023,Beenakker_2023,Haegeman:2024qgf,Zakharov:2024hlo,Zakharov:2024xcg}. However, as the author mentioned in \cite{Fukusumi:2022xxe}, we stress the significance of the studies on the corresponding chiral model in the system with boundary\cite{Li_2008,Qi_2012}, because of its close connection to boson-fermion correspondence\cite{Skyrme:1961vq,Coleman:1974bu,Mandelstam:1975hb} or Rogers-Ramanujan type identities\cite{Andrews:1984af,Campbell_2024} (or open string field theory).}. Hence it is natural to expect there exists a CFT analog of similarity transformation, but the framework is still under development. In this section, we propose a possible unified framework of local non-Hermitian systems (or the corresponding non-local Hermitian systems) based on the arguments of the bulk-edge correspondence in the previous sections. The interpretation of $3d-3d$ correspondence in the literature leads to a nontrivial series of models by combining the studies of nonunitary CFTs and their integrable flows\cite{Gannon:2003de,Castro-Alvaredo:2017udm,Kikuchi:2022rco,Fukusumi_2022,Fukusumi_2022_c,Lencses:2022ira,Klebanov:2022syt,amslaurea11308,Lencses:2023evr,Lencses:2024wib,Tanaka:2024igj}.

\begin{figure}[htbp]
\begin{center}
\includegraphics[width=0.5\textwidth]{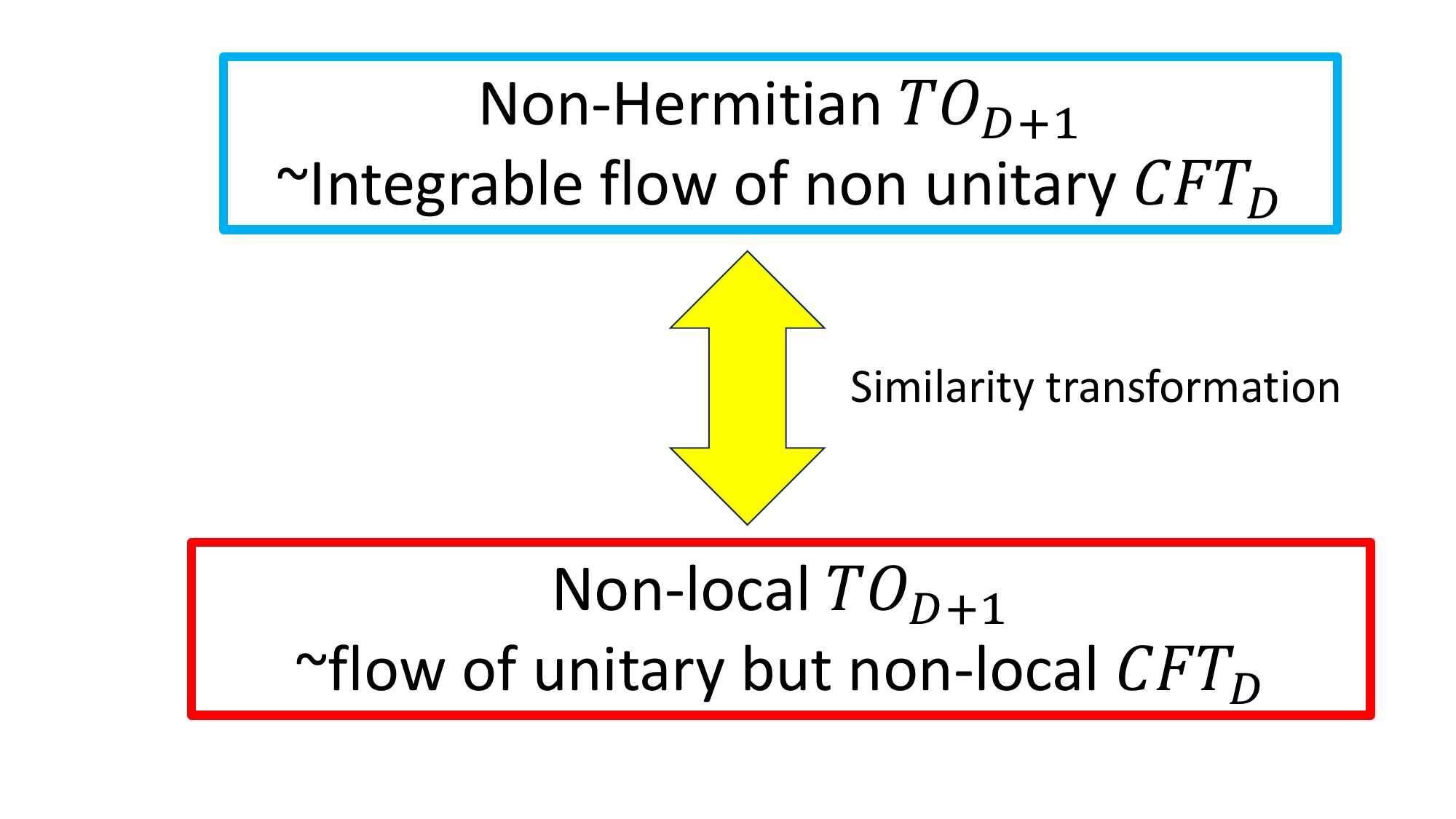}
\caption{Implication of similarity transformation in the non-Hermitian many-body systems and similarity duality of different CFTs. This corresponds to a generalization of arguments in \cite{Read:2007cv} for nonlocal or non-Hermitian models.}
\label{similarity_duality}
\end{center}
\end{figure}

A typical example of the operation corresponding to the above similarity transformation can be seen in the studies on the correspondence between Symplectic fermion and Dirac fermion\cite{Guruswamy:1996rk,Hsieh:2022hgi,Ryu:2022ffg}.
In this type of correspondence, whereas the theories share the same Hilbert space or quantum states, the operators constructing the correlation functions can be different and one can interpret this as a kind of level-rank duality in the literature\footnote{However, this condition is slightly different from conventional level-rank duality because the effective central charge can be different, for example in $SU(N)_{K}$-$SU(K)_{N}$ type duality. In this duality, one can map (only map) the quantum states, but the resultant fusion rule of operators becomes the same.}. More recently, a correspondence between $Sp(2N)_{k-1/2}$ minimal model and $Osp(1,N)_{k}$ Wess-Zumino-Witten model has been proposed in \cite{Creutzig:2024ljv} in the study of the  $3d-3d$ correspondence in TQFTs\cite{Cho:2022kzf,Gang:2023ggt,Gang:2023rei,Baek:2024tuo,Gang:2024tlp}. We claim that this type of correspondence is interpreted as a generalization of the correspondence between Symplectic fermion and Dirac fermion studied in the literature. For the latter use, we call the general correspondence between unitary and nonunitary models as ``\emph{similarity duality}"(FIG\ref{similarity_duality}), and we call this particular type of correspondence as \emph{dimensional irreduction} from its close connection to the notion of dimensional reduction as we demonstrate in \ref{dimensional_irreduction_section}.

 The dimensional irreduction is including the correspondence between the $M(2,2N+3)$ minimal model and the $Osp(1,2)_{N}$ WZW model by fixing $k=1$. The similarity of these two models can be seen in \cite{Wood:2018xqk} and one can also see related coset models and the corresponding lattice model\cite{Saleur:1989gj,Bershadsky:1989tc,Saleur:2001cw,Saleur:2003zm,Yang:2008vb}. Moreover, the unitary minimal model can be described by the $Sp(N)$ coset model\cite{Goddard:1984vk,Goddard:1985jp,Goddard:1984hg}, and this can be thought of as a generalization. We remind again of the correspondence of quantum states between Dirac fermion and symplectic fermion \cite{Guruswamy:1996rk,ludwig2000freefieldrepresentationosp22,LeClair:2007aj,Creutzig:2008an}, which has a close connection to the models we mentioned. Hence, we support the proposal in \cite{Creutzig:2024ljv}. For comparison, we note the effective and exact conformal dimensions of both theories as,
\begin{equation}
h_{1,s}^{\text{eff}}=h_{1,s}^{M(2,2N+3)}+\frac{N(N+2)}{2(2N+3)}=\frac{J(2J+1)}{2N+3}=h^{Osp}_{J=\frac{N+1-s}{2}}
\end{equation}
where $h_{1,s}^{M(2,2N+3)}$ is the conformal dimension of minimal model\cite{Belavin:1984vu} labeled by the Kac indeces $(1,s)$ with $s=1,..,N+1$ and   $h^{Osp}_{J}$ is conformal dimension of spin $J$ primary fields with $J=0,1/2,1,...,N/2$ as in \cite{Ennes:1997vt}.
The effective central charge of both theories is $c=2N/(2N+3)$ \cite{Ferrari:2023fez}\footnote{Interestingly, the twice of this effective central charge is the same as that of $Z_{2N+1}$ parafermion. Moreover, the product theory of $M(2,5)\times M(2,5)$ is the same as $M(3,10)$\cite{Quella:2006de}. The readers can easily check the effective central charge of the minimal model by the following formula $c^{\text{eff}}=1-(6/pq)$. In general, there may exist some nontrivial relationship between different CFTs with the same $p\times q$ (probably exact marginal deformations between them), but this is a future problem. Similar correspondeces can be seen in \cite{Dunning:2002cu}.}.

\begin{figure}[htbp]
\begin{center}
\includegraphics[width=0.5\textwidth]{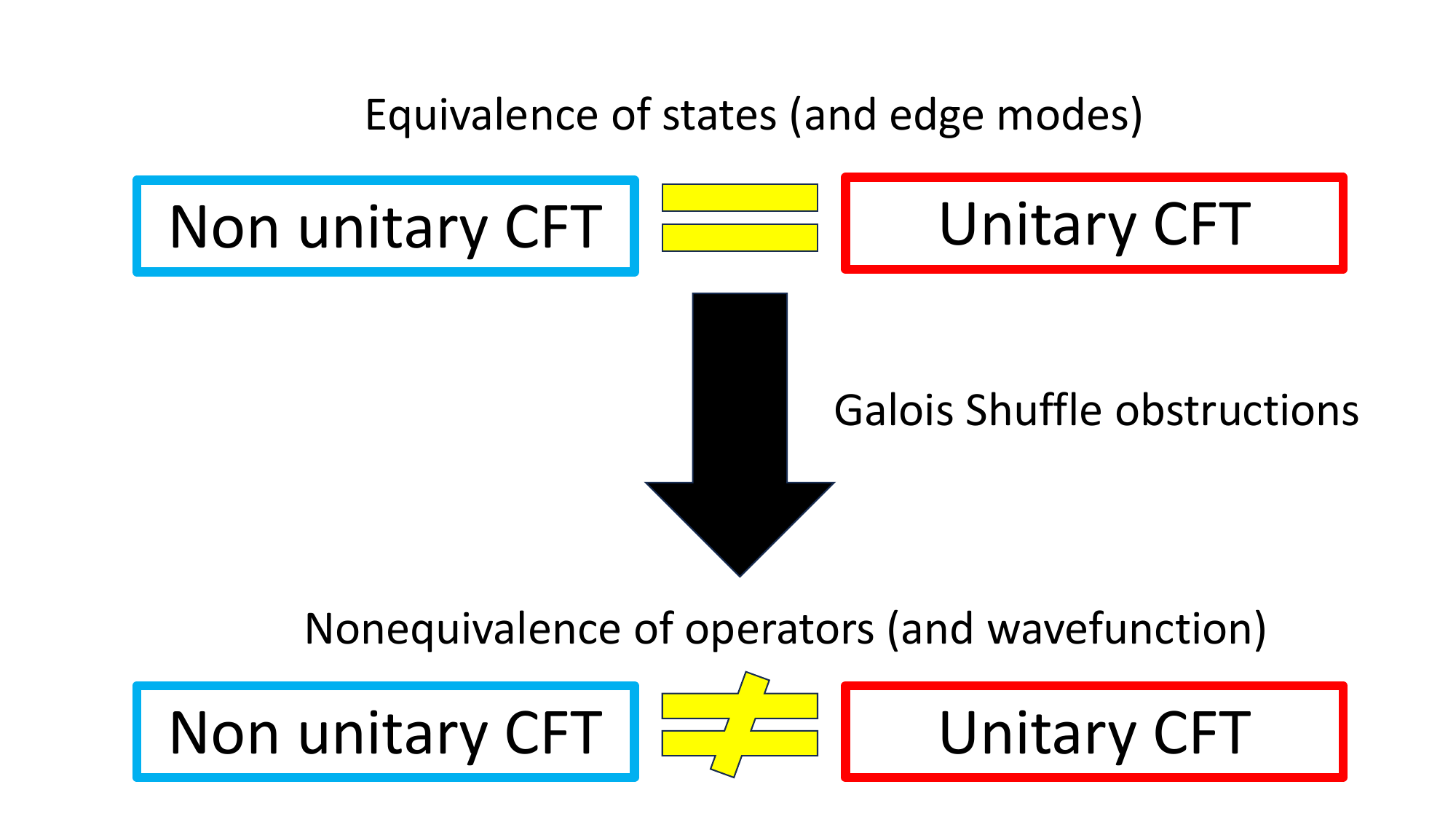}
\caption{Galois shuffle and obstruction of the correspondence of the operators in similarity dual models. Related observations can be seen in the literature on gapless FQHEs \cite{Guruswamy:1996rk,Fukusumi_2022,Fukusumi_2022_c,Milovanovic:1996nj,Ino:1998by}. The mismatch of operators results in different wavefunctions of $3d$ TOs by bulk-edge correspondence and different $4d$ theories by uplifting the models\cite{Parisi:1979ka}. Related discussions can be seen in \cite{Dedushenko:2023cvd}. When studying their symmetries or defects, it may depend on the application of the Verlinde formula based on their effective vacuum corresponding to their states or their exact vacuums corresponding to their operators.}
\label{Galois_shuffle_symmetry}
\end{center}
\end{figure}

More intuitively, one can understand this phenomenon as a consequence of the boson-fermion correspondence\cite{Skyrme:1961vq,Coleman:1974bu,Mandelstam:1975hb}. The following Rogers-Ramanujan identity for the $M(2, 2N+3)$ model, or Andrew-Gordon identity, is widely known in the literature\cite{Nahm:1992sx},
\begin{equation} 
\chi_{h_{1,s}}=q^{h_{1,s}-c/24}\sum_{n_{1},...,n_{N}\ge 0}\frac{q^{\sum_{i}m_{i}^{2}}}{\prod_{i}(q)_{n_{i}}}
\label{Andrew-Gordon_eq}
\end{equation}
where $m_{i}$ is the summation $n_{i}+...+n_{N}$ and $(q)_{n}$ is a product $(1-q)\times ...\times (1-q^{n})$. The summation appearing in Eq.\eqref{Andrew-Gordon_eq} can be considered as a fermionic sum\cite{Blondeau-Fournier:2017otv,Blondeau-Fournier:2010fkq,Welsh:2002jq}.
Because of the close connection between the boson-fermion correspondence and Rogers-Ramanujan identity\cite{Andrews:1984af,Campbell_2024}, one may expect the appearance of $Z_{2}$ simple current corresponding to the fermionic representation for $M(2,2N+3)$ based on the recent analysis of the fermionization\cite{Runkel:2020zgg,Hsieh:2020uwb,Runkel:2022fzi}. However, it does not appear even in the simplest minimal model, $M(2,5)$. Hence the fermionic summation should correspond to an unconventional fermionic model outside of the simple current extension\cite{Gato-Rivera:1991bqv,Gato-Rivera:1991bqv,Gato-Rivera:1991bcq}, and this corresponds to the $Osp(1,2)_{N}$ WZW model. In other words, by applying the analogy with the correspondence between the symplectic fermion and Dirac fermion, one can interpret the $M(2,2N+3)$ minimal model and corresponding integrable flow as hidden supersymmetry-protected RG flow, which can be thought of as a generalization of the flow \cite{Furuya:2015coa,Kikuchi:2019ytf,Numasawa:2017crf,Fukusumi_2022_c,Kikuchi:2022ipr}. However, for future research, we emphasize that the structure of the operators in the dimensional irreduction can be different whereas their Hilbert spaces are the same. This kind of mismatch of the operators can be seen in the correspondence between Dirac fermion and symplectic fermion \cite{Guruswamy:1996rk} \footnote{Whereas the correspondence of the operators seems to be conjectured in \cite{Creutzig:2024ljv}.}. We also conjecture the symmetry of these similarity dual models is determined by the modular $S$ property respecting quantum states or operators(FIG. \ref{Galois_shuffle_symmetry}).

This argument on quantum states and operators plays a fundamental role in considering the corresponding fractional quantum effect because the structure of the wavefunction of FQHE is determined by the $Z_{N}$ symmetry charge of the operator. Hence it is impossible to obtain the modular $S$ invariant partition function from a class of nonunitary CFTs in the conventional way because the modular property is governed by $Z_{N}$ charge of quantum states. The mismatch of the charges of states and operators is a consequence of Galois shuffle\cite{Gannon:2003de} and this can play a fundamental role in the classification of non-unitary CFTs and the corresponding topological orders\cite{Harvey:2019qzs,Fukusumi_2022,Fukusumi_2022_c}. It is also worth noting that the structure of Galois shuffle can be defined in the $M(2,2N+3)$ model whereas the theory has no $Z_{2}$ fields, by considering the property of modular $S$ or property of quantum states and topological symmetry operators. This can also be understood as a consequence of the hidden $Z_{2}$ or supersymmetry, and one can call this type of Galois shuffle a hidden Galois shuffle from hidden supersymmetry.

\begin{figure}[htbp]
\begin{center}
\includegraphics[width=0.5\textwidth]{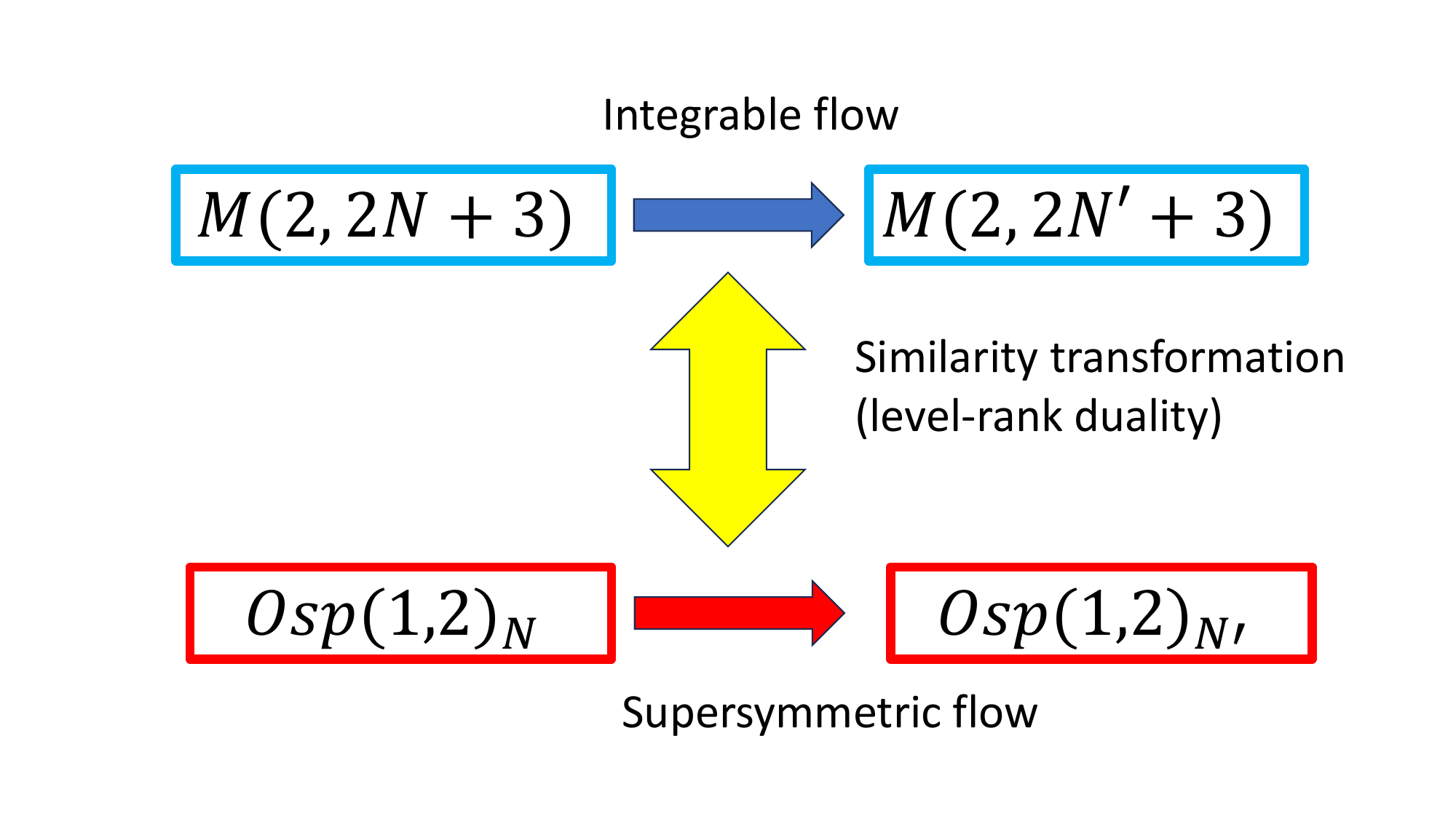}
\caption{Implication of similarity transformation in the flow of Yang-Lee CFTs and the corresponding $Osp(1,2)$ WZW model. The relation to integrable flow and supersymmetry can be seen in a wide literature\cite{Gaiotto:2012np,Kikuchi:2022ipr,Fukusumi_2022_c}.}
\label{hidden_supersymmetry}
\end{center}
\end{figure}

\subsection{dimensional irreduction}
\label{dimensional_irreduction_section}

The idea of boson-fermion correspondence and its symplectic fermion representation can be seen in the notable work on the Goddard-Kent-Olive coset construction or ``quark model" (it might be surprising for the readers, the original coset representation of the unitary minimal model \cite{Goddard:1984vk,Goddard:1985jp,Goddard:1984hg} has been studied in the fermionic representation. We also note the legacy of David Olive \cite{https://doi.org/10.48550/arxiv.2009.05849} for a historical aspect). Because of its close connection to string theories or supersymmetric models, one may expect some kind of arguments related to this direction. One of the most notable examples is the dimensional reduction relating $D-2$ dimensional Yang-Lee CFT and $D$ dimensional supersymmetric model\cite{Parisi:1979ka}. However, whereas the uplift of $D-2$ dimensional CFT to $D$ dimensional CFT has been proposed recently in\cite{Trevisani:2024djr}, the original proposal of the dimensional reduction suffers from several problems, especially in lower $D$\cite{Imry:1975zz,Fytas:2016itl,Fytas:2019zdk}. The dimensional irreduction and the related $3d-3d$ correspondence can be thought of as such exceptional theory in this paradigm, and this indicates the hidden supersymmetry for Yang-Lee CFT is fundamental(FIG. \ref{hidden_supersymmetry}). In this work, we have concentrated our attention on supersymmetric cases, but the generalization to $Z_{N}$ models or fractional supersymmetric models seems to be possible by revisiting various level-rank duality and coset conformal field theory. In some class of models, for example, $Sp(2N)_{K}\times Sp(2K)_{N}=SO(4NK)_{1}$ level-rank duality, the cancelation and matching of Galois shuffle may play a fundamental role by interpreting the lefthand side as symplectic fermions. By choosing $N=K=1$, this corresponds to the model in \cite{Hsieh:2022hgi}. \footnote{Interestingly, the central charge of the $SO(4M)$ model, $2M$, changes by $2$ depending on the integer $2$.}

\begin{figure}[htbp]
\begin{center}
\includegraphics[width=0.5\textwidth]{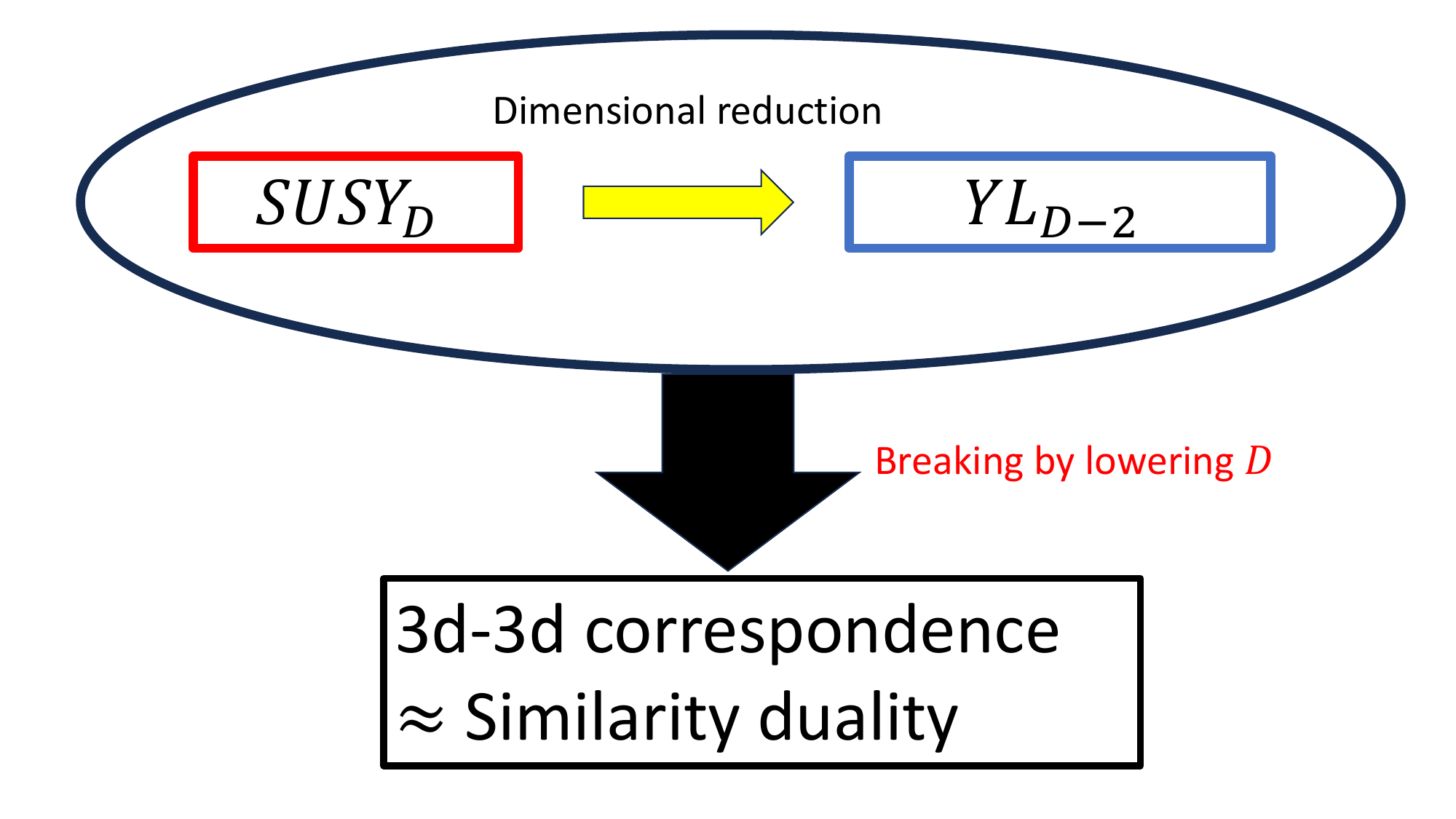}
\caption{A picture of dimensional irreduction as breaking of dimensional reduction. $SUSY_{D}$ denotes a $D$ dimensional supersymmetric quantum field theory and $YL_{D-2}$ denotes the corresponding $D-2$ dimensional Yang-Lee CFT. A class of $3d-3d$ correspondence can be thought of as breaking of yellow arrow by lowering the space dimension $D$.}
\label{hidden_supersymmetry}
\end{center}
\end{figure}

Before going to the concluding remark, we note two other types of correspondence between CFTs at the level of the operator. First, it is known that the CFTs with different central charges can have the same set of conformal conformal dimensions\cite{Batchelor:1994zz}. Moreover, the corresponding TO has been proposed in \cite{Moharramipour:2023dmh}. Generalized symmetry analysis of these models might be suggestive because the conformal dimensions do not appear in this kind of analysis in usual. Second, a mysterious correspondence of fusion rules of conformal field theories has been proposed by using the Hecke operation (or symmetry orbifolding)\cite{Harvey:2019qzs}. Surprisingly, they have shown that the fusion rule of CFTs constructed from a non-unitary CFT can be preserved under a condition.

\section{Conclusion}
\label{conclusion}

In this work, we have reformulated the structures of edge modes of TO in the language of integrable RG flow. In our formalism, one can understand the existing phenomena in a unified way and the conciseness of our formalism is a strong benefit to study further nontrivial phenomena. As an open problem, we remark on the fundamental significance of studying the correspondence of the following three objects: operators, states, and symmetries. In the recent findings, it becomes gradually evident that the correspondence of these three objects can be broken in a rigorous sense at the level of quantum field theories or the realization of many-body systems. However, as the author has clarified in \cite{Fukusumi:2023psx,Fukusumi:2024cnl}, modified versions of this correspondence (or algebraic structure) can exist. Hence, as we have discussed in this work, further studies on the various structures in QFTs and their mutual relations or consistencies are fundamentally important. In other words, it is necessary to understand RG as completely algebraic objects. This view is fundamentally important to formulate RGs as rigorous or mathematical objects and related approaches can be seen in \cite{Crnkovic:1989ug,Quella:2006de,Gaiotto:2012np,Klos:2019axh,Klos:2021gab,Zeev:2022cnv,Kikuchi:2022rco,Kikuchi:2022ipr,Cogburn:2023xzw}. 

As a typical example, we note the hidden (fractional) supersymmetric interpretation of Rogers-Ramanujan-type identities which have close connections to various topics in contemporary mathematics. One can also consider a generalization of our argument to the systems in higher space-time dimensions because except for Section \ref{non-Hermitian_correspondence}, we have not used the special property of $2+1$ dimensional topological orders. Hence, as the author has remarked in \cite{Fukusumi:2023psx,Fukusumi:2024cnl}, studies on integrable deformations on ancillary CFTs \cite{Nishioka:2022ook}, which are higher dimensional analogs of the chiral CFTs, are fundamentally important for a unified understanding of TOs. We also note several existing literature on the modular property of higher dimensional models\cite{Shaghoulian:2015kta,Shaghoulian:2015lcn,Chen_2016,Park:2016xhc,Belin:2016yll,Chen_2017,Luo:2022tqy,Lei:2024oij,Hofman:2024oze} for further studies.

\section{Acknowledgement}

We thank Bo Yang and Guangyue Ji for the past collaboration related to this project and Tan Fei and Yuzhu Wang for helpful discussions on the nonlinear dynamics of fractional quantum Hall fluid. We thank Ken Kikuchi for the discussions on the fermionization and indications on non-unitary models and Takamasa Ando for sharing his knowledge on symmetry topological field theory. We thank Po-Yao Chang and Chang-Tse Hsieh for the helpful discussions on the non-Hermitian systems and Yutaka Matsuo for many helpful comments on the level-rank duality and the historical aspects of string compactification and Himanshu Parihar for the discussion on the Lie group symmetry in AdS/CFT. It is a great pleasure to thank Hosho Katsura for sharing a lot of references and ideas in various fields in this work. Finally, we thank Duncan Haldane for the questions and comments on the implications of the modular properties in topologically ordered systems and the related discussion of the dynamical properties of chiral edge modes, which are the central subjects in this work. We acknowledge the support from NCTS and CTC.

\appendix
\section{Bulk property of mixed state topological order by gluing technique}
\begin{figure}[htbp]
\begin{center}
\includegraphics[width=0.5\textwidth]{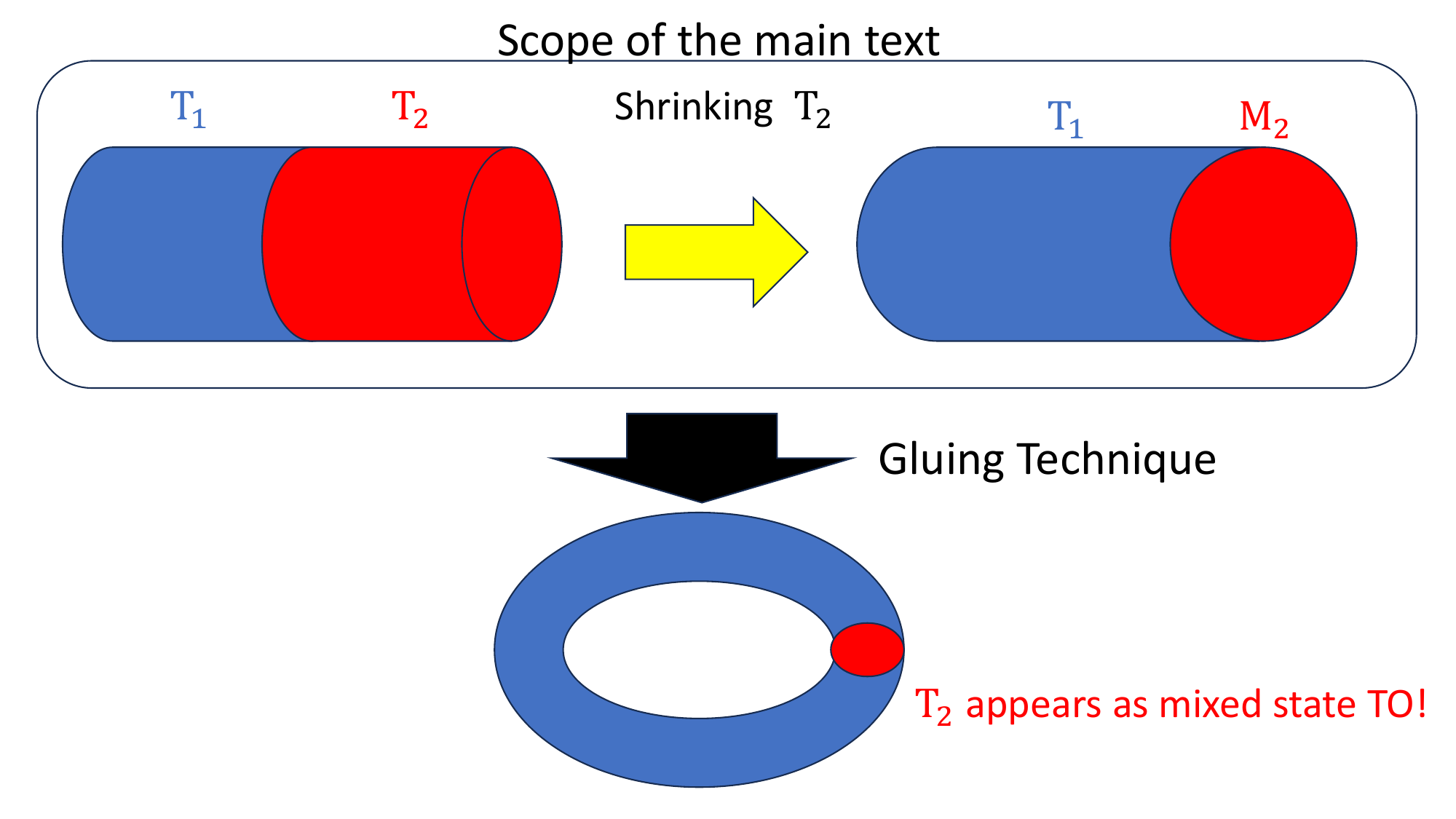}
\caption{Relationship between the arguments in the main text and those on the mixed state TOs. Related arguments can be seen in \cite{Fukusumi:2022xxe,Fukusumi_2022_c,Fukusumi:2024cnl,Kikuchi:2024ibt}.}
\label{mixed_TO}
\end{center}
\end{figure}

Recently, it has been proposed that the TO of mixed states can have a nontrivial structure compared with the ground states\cite{Fan:2023rvp,You:2024mth,Lessa:2024gcw,Sala:2024ply}. One of the nontrivial points in this phenomenon is the appearance of $\mathcal{M}_{2}$ or $\mathcal{T}_{2}$ in the main text which is difficult to predict only from the ground state bulk TO $\mathcal{M}_{1}$ or $\mathcal{T}_{1}$\cite{Wang:2023uoj,Sohal:2024qvq,Ellison:2024svg,Kikuchi:2024ibt}. However, as we have demonstrated in the main text, by considering the system with boundaries, one can explain the analogous phenomena in the scope of existing formalisms. Moreover, the phenomena in mixed state TOs seem to be consistent with the gluing technique of TQFTs\cite{Wen:1990se,Qi_2012}(FIG. \ref{mixed_TO}). Further investigations of the many-body systems seem necessary, but it should be kept in mind that the phenomena themselves are in the scope of usual bulk-edge correspondence (combined with RG analysis or anyon condensation)\cite{Fukusumi:2024cnl}. For this purpose, we remark that studying a chiral or ancillary CFT is fundamental to constructing the corresponding TOs.\cite{Fuchs:2002cm,Fuchs:2004dz,Fukusumi:2023psx,Fukusumi:2024cnl}.

\bibliography{emergent}

\end{document}